\documentclass[%
 reprint,
 amsmath,amssymb,
 aps,
]{revtex4-1}
\usepackage{amsmath}
\usepackage{graphicx}
\usepackage{dcolumn}
\usepackage{bm}

\usepackage{rotating}
\usepackage{tabularx}
\usepackage{makecell}
\usepackage{booktabs}
\usepackage{colortbl}
\usepackage[table]{xcolor}
\usepackage{array}
\newcommand {\pa}       {\rm PYTHIA}
\newcommand {\spp}      {\sqrt{\rm s}}
\newcommand {\sauau}  {\sqrt{\rm s_{\rm NN}}}
\newcommand {\rpp}      {R_{pp}}
\newcommand {\pT}       {p_{T}}
\newcommand {\pTM}    {\langle{p_{T}}\rangle}
\newcommand {\Nch}     {N_{ch}}
\newcommand {\NchM}  {\langle{N_{ch}}\rangle}
\newcommand {\pipi}      {\pi^-/\pi^+}
\newcommand {\kk}        {K^-/K^+}
\newcommand {\pp}        {\overline{p}/p}
\newcommand {\kpipos}  {K^+/\pi^+}
\newcommand {\kpineg}  {K^-/\pi^-}
\newcommand {\ppipos}  {p/\pi^+}
\newcommand {\ppineg}  {\overline{p}/\pi^-}
\newcommand{\Lks}        {\Lambda/K^{0}_{s}}
\newcounter{fofo}

\begin{document}

\title{Multiplicity Dependence of Charged Particle, $\phi$ Meson and Multi-strange Particle Productions in p+p Collisions at $\spp$ = 200 GeV with PYTHIA Simulation}

\author{Shenghui Zhang}
\author{Long Zhou}
\author{Yifei Zhang}
\author{Mingwei Zhang}
\author{Cheng Li}
\author{Ming Shao}
\author{Yongjie Sun}
\author{Zebo Tang}
\affiliation{University of Science and Technology of China, Hefei 230026, China}
\date{\today}
\begin{abstract}
We report the multiplicity dependence of charged particle productions for $\pi^{\pm}$, $K^{\pm}$, $p$, $\overline{p}$ and $\phi$ meson at $|y| < 1.0$ in p+p collisions at $\spp$ = 200 GeV with $\pa$ simulation. The impact of parton multiple interactions and gluon contributions is studied and found to be possible sources of the particle yields splitting as a function of $\pT$ with respect to multiplicity. No obvious particle species dependence for the splitting is observed. The multiplicity dependence on ratios of $\kpineg$, $\kpipos$, $\ppineg$, $\ppipos$ and $\Lks$ in mid-rapidity in p+p collisions is found following the similar tendency as that in Au+Au collisions at $\sqrt{s_{NN}}$ = 200 GeV from RHIC, which heralds the similar underlying initial production mechanisms despite the differences in the initial colliding systems.

\begin{description}
\item[PACS number]
25.75.Cj
\end{description}

\pacs{25.75.Cj}
\end{abstract}
\maketitle

\section{\label{sec:level1}Introduction}

Searching for a novel form of nuclear matter with deconfined quarks and gluons created in ultra-relativistic heavy-ion collisions is the main goal of high energy nuclear physics. This strongly coupled matter, so called Quark-Gluon Plasma (sQGP) ~\cite{QGP, QGP1, QGP2} emerges its properties by experimentally comparing with elementary particle collisions. Measurements of particle production in proton-proton (p+p) collisions are critical to provide a baseline for understanding the interactions in the QGP created in heavy-ion collisions. Since past a decade, there have been many existing measurements on the multiplicity dependence or centrality dependence of particle productions in heavy-ion collisions ~\cite{STARpikpFUQIANG, STARpikp, STARstrangeness, ALICEmult-strangeness}. In particular, the production mechanism of hadrons contain strangeness is believed as a signature for QGP formation in heavy-ion collisions ~\cite{Theory-Strangeness-QGP1, Theory-Strangeness-QGP2, Theory-Strangeness-QGP3}. To support this, it is worthy of studying the multiplicity dependence of strangeness production in elementary particle collisions without any medium effect. Recently, the strangeness enhancement in high-multiplicity p+p collisions was observed by ALICE experiment ~\cite{ALICENature}, which is similar to those observed in the heavy-ion collisions ~\cite{ALICEPdPd}, where a hot-dense medium is created. This is in contrast to our knowledge that the elementary particle collisions create cold and tiny system, thus drawing a lot of interest. However, the p+p colliding system is always treated as fundamental particle collisions and the particle production with respect to multiplicity is barely studied, especially at a colliding energy of a few hundred GeV at Relativistic Heavy Ion Collider (RHIC). Thus, it is also of interest to investigate the multiplicity dependence of particle production in p+p collisions at RHIC energy to see if there is any similarity as in Au+Au collisions can be found. 

In this paper, we report the multiplicity dependence of particle productions for $\pi^{\pm}$, $K^{\pm}$, $p$ and $\overline{p}$ in p+p collisions at $\spp$ = 200 GeV based on $\pa$ simulation. We study the multiplicity dependence of particle yields as a function of transverse momentum ($\pT$) for $\pi^{\pm}$, $K^{\pm}$, $p$, $\overline{p}$ and $\phi$ and ratio of $\kpineg$, $\kpipos$, $\ppineg$ and $\ppipos$. There are three important sources of the $\pT$ distribution splitting with multiplicity including jet fragmentation, parton multiple interactions and gluon contributions. Since the jet fragmentation effect has been discussed in Ref. ~\cite{XINNIAN-jet}, we focus on the effect of parton multiple interactions and gluon contributions. The particle ratios of $\kpineg$, $\kpipos$, $\ppineg$ and $\ppipos$ with respect to multiplicity and a comparison to that in Au+Au collisions at $\sauau$ = 200 GeV measured from RHIC-STAR ~\cite{STARpikpFUQIANG, STARpikp} are presented. The $\Lks$ ratios in p+p collisions at $\spp$ = 200 GeV in RHIC energies are also presented and compared with experimental results in Au+Au collisions.

The paper is organized as follows: Section II presents the simulation process and detailed $\pa$ settings. Simulation results on charged particle and $\phi$ meson $\pT$ spectra, average $\pTM$, particle yield ratios and related discussions are presented in Section III. In the end, Section IV gives the summary. 

\section{\label{sec:level1}SIMULATION PROCESSES}
\begin{figure}[h]
\centering
\includegraphics[width=0.45\textwidth]{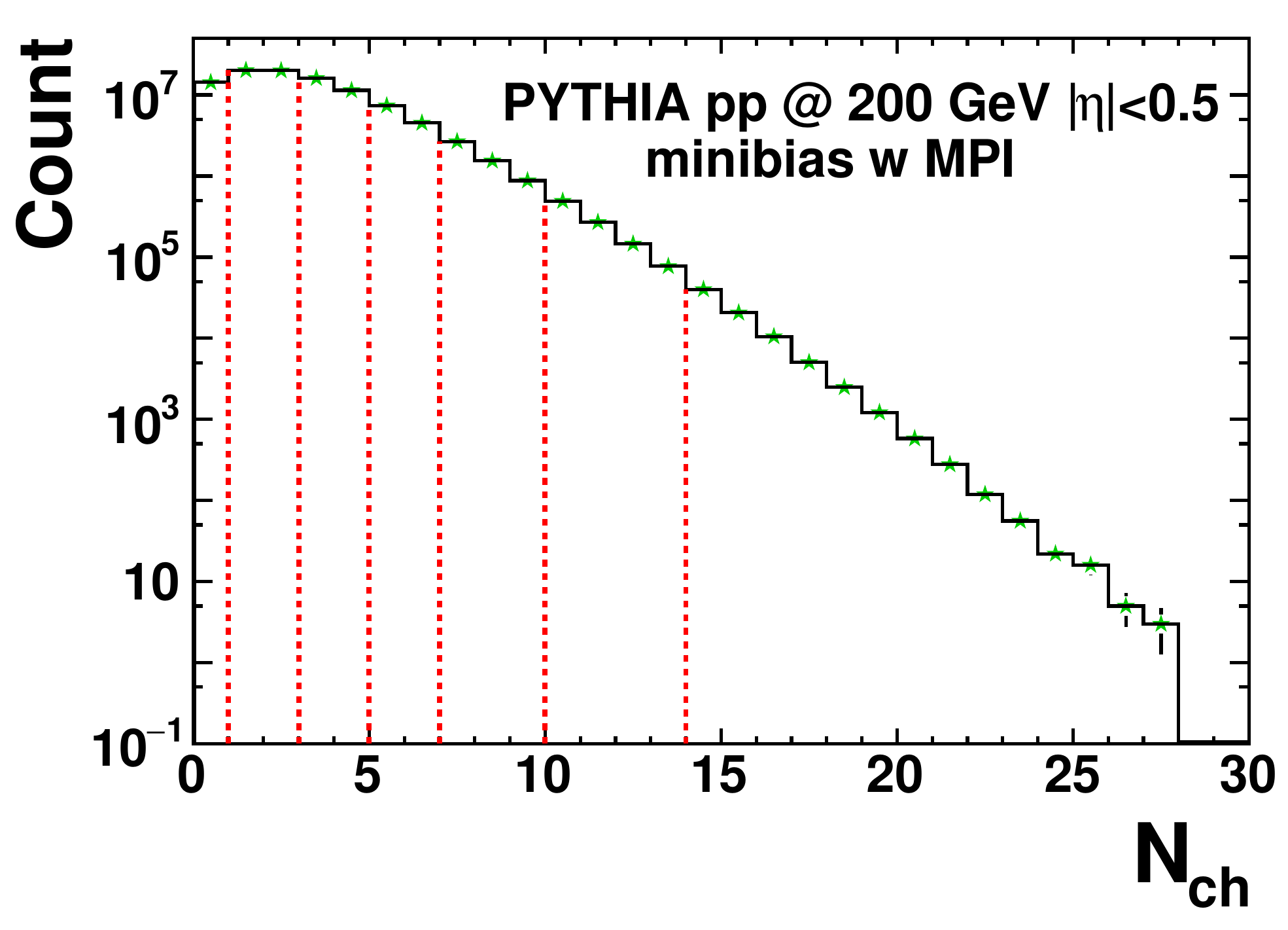}
\caption{\label{fig1} Charged particle multiplicity distribution obtained from $\pa$ within $|\eta| < 0.5$ in p+p collisions at 200 GeV. The regions between dotted line indicate the multiplicity bins $\NchM$ used in the analysis.}
\end{figure}

The $\pa$ program is widely used for event generation in high-energy physics, to simulate multiparticle production in collisions between elementary particles ~\cite{PTHIA}. In this work, the $\pa$ version used is 6.416. The p+p collision events at $\spp$ = 200 GeV are generated and categorized in 6 multiplicity bins as shown in Fig.~\ref{fig1} to study the multiplicity dependence of $\pi^{\pm}$, $K^{\pm}$, $p$, $\overline{p}$ and $\phi$ productions. The initial configurations in $\pa$ are as follows:

MSEL(1) include: ISUB = 11 $q_{i}q_{j} \to q_{i}q_{j}$, 12 $q_{i}\overline{q}_{i} \to q_{k}\overline{q}_{k}$, 13 $q_{i}\overline{q}_{i} \to gg$, 28 $q_{i}g \to q_{i}g$, 53 $gg \to q_{k}\overline{q}_{k}$, 68 $gg \to gg$, 96 semihard QCD $2 \to 2$. If all processes are c-

\begin{table*}[htbp]
\caption{$\NchM$ in different multiplicity for different mechanisms.}
\centering
\begin{tabular}{p{80pt}<{\centering}p{40pt}<{\centering}p{40pt}<{\centering}p{40pt}<{\centering}p{40pt}<{\centering}p{40pt}<{\centering}p{40pt}<{\centering}p{68pt}<{\centering}}
\hline
\hline
Mechanisms&$\langle{N_{ch1}}\rangle$&$\langle{N_{ch2}}\rangle$&$\langle{N_{ch3}}\rangle$&$\langle{N_{ch4}}\rangle$&$\langle{N_{ch5}}\rangle$&$\langle{N_{ch6}}\rangle$&$\langle{N_{ch}(minibias)}\rangle$ \\
\hline
minibias w MPI&1.50&3.41&5.38&7.65&10.80&14.99&3.14\\
minibias wo MPI&1.48&3.38&5.31&7.45&10.52&14.84&2.55\\
gluon off w MPI&1.51&3.40&5.36&7.62&10.77&14.95&2.99\\
gluon off wo MPI&1.46&3.34&5.27&7.35&10.42&15.04&2.28\\
\hline
\hline
\end{tabular}
\end{table*}
\noindent  onsidered, called minimum bias (minibias) event. The gluon contribution is studied via changing the ISUB 28, 53 and 68 processes setup.

MSTP(81, 1): turn multiple interaction on.

MSTP(61, 1), MSTP(71, 1): initial- and final-state QCD radiation is added to the above processes.

MSTP(51, 7): CTEQ5L parton distribution function.

MSTP(33, 1): a common K factor is used, as stored in PARP(31).

PARP(31, 1.5): (D = 1.5) common K factor multiplying the differential cross section for hard parton-parton processes.


\section{\label{sec:level1}Results and Discussions}
\subsection{\label{sec:level2}Ratio of particle production yields}
The p+p collision events generated from $\pa$ produce charged particle ($\pi^{\pm}$, $K^{\pm}$, $p$, $\overline{p}$, $e^{\pm}$ and $\mu^{\pm}$) within $|\eta| < 0.5$ at 200 GeV. The number of them is called charge multiplicity, $\Nch$. The multiplicity distributions are shown in Fig. 1. Six multiplicity bins are chosen according to the equal number of events in each multiplicity bin. The events without any multiplicity selection is called minibias events. The ratio of the particle production yield in each multiplicity bin over that in minbias events and normalized by $\NchM$ (the mean of $\Nch$) are used to study the event activities with respect to different multiplicities, which can be defined as:$$\rpp = \frac{dN/d\pT(mult, \pT)/\langle{\Nch(mult)}\rangle}{dN/dp_{T}(minibias, \pT)/\langle{\Nch(minibias)}\rangle} , (1)$$ 
\begin{figure}[h]
\centering
\includegraphics[width=0.5\textwidth]{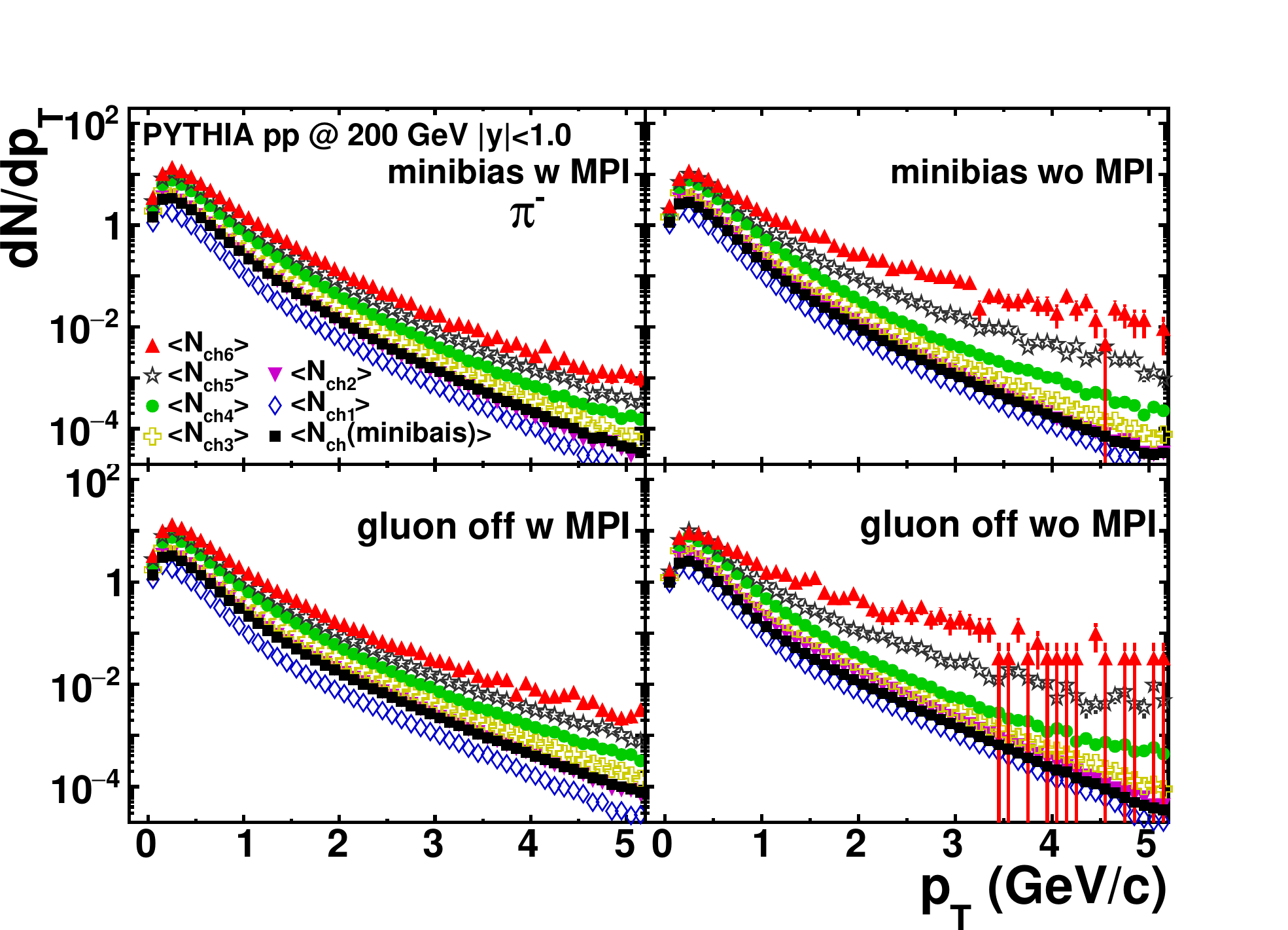}
\caption{\label{label}$\pT$ spectrum of $\pi^{+}$ in different multiplicity bins with and without parton multiple interactions and gluon contributions in p+p collisions at 200 GeV.}
\end{figure}
\begin{figure}[h]
\centering
\includegraphics[width=0.5\textwidth]{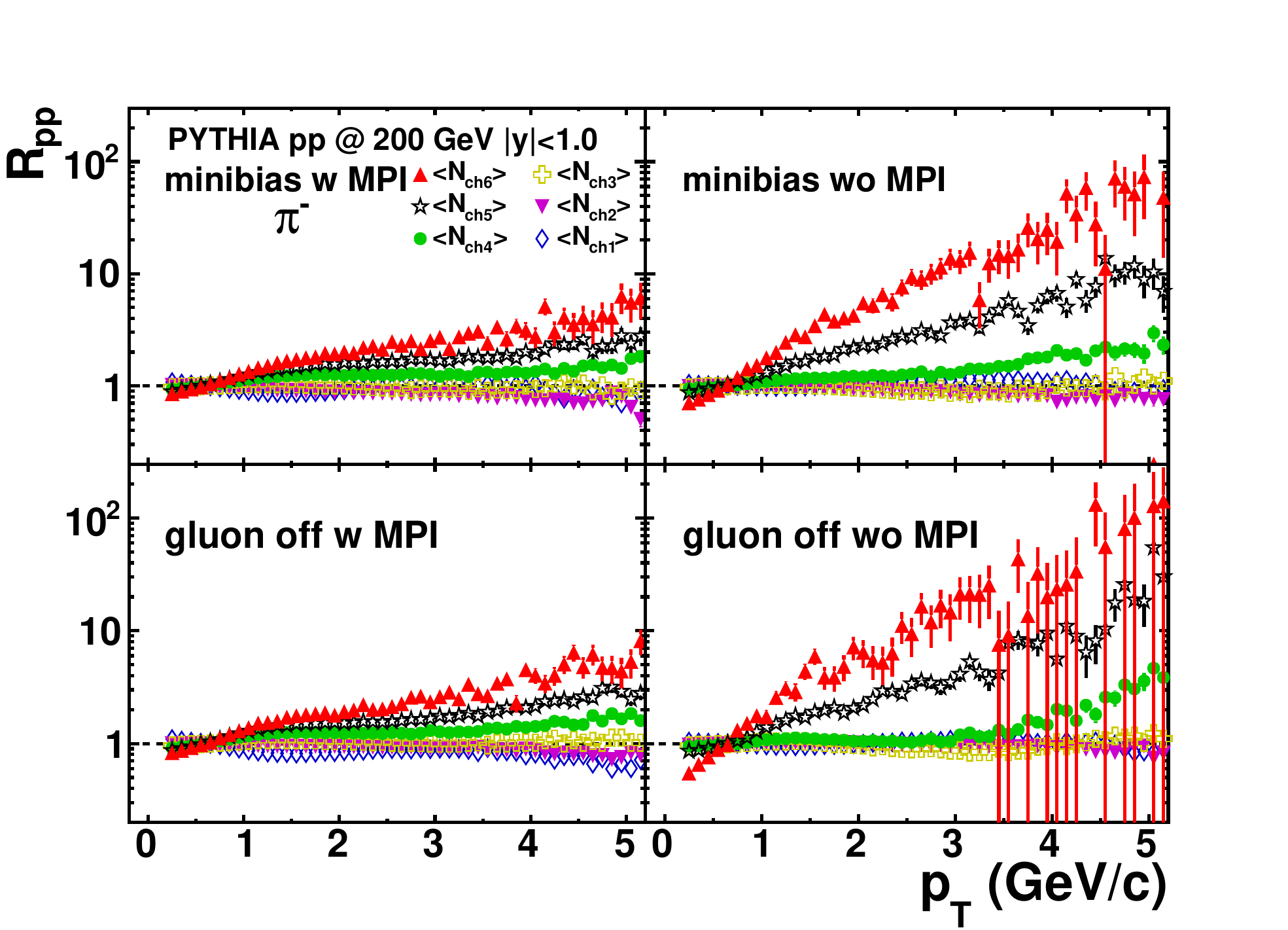}
\caption{\label{label}$\rpp$ spectrum as a function of $\pT$ for $\pi^{-}$ in different multiplicity bins with and without parton multiple interactions and gluon contributions in p+p collisions at 200 GeV.}
\end{figure}
\begin{figure}
\includegraphics[width=0.5\textwidth]{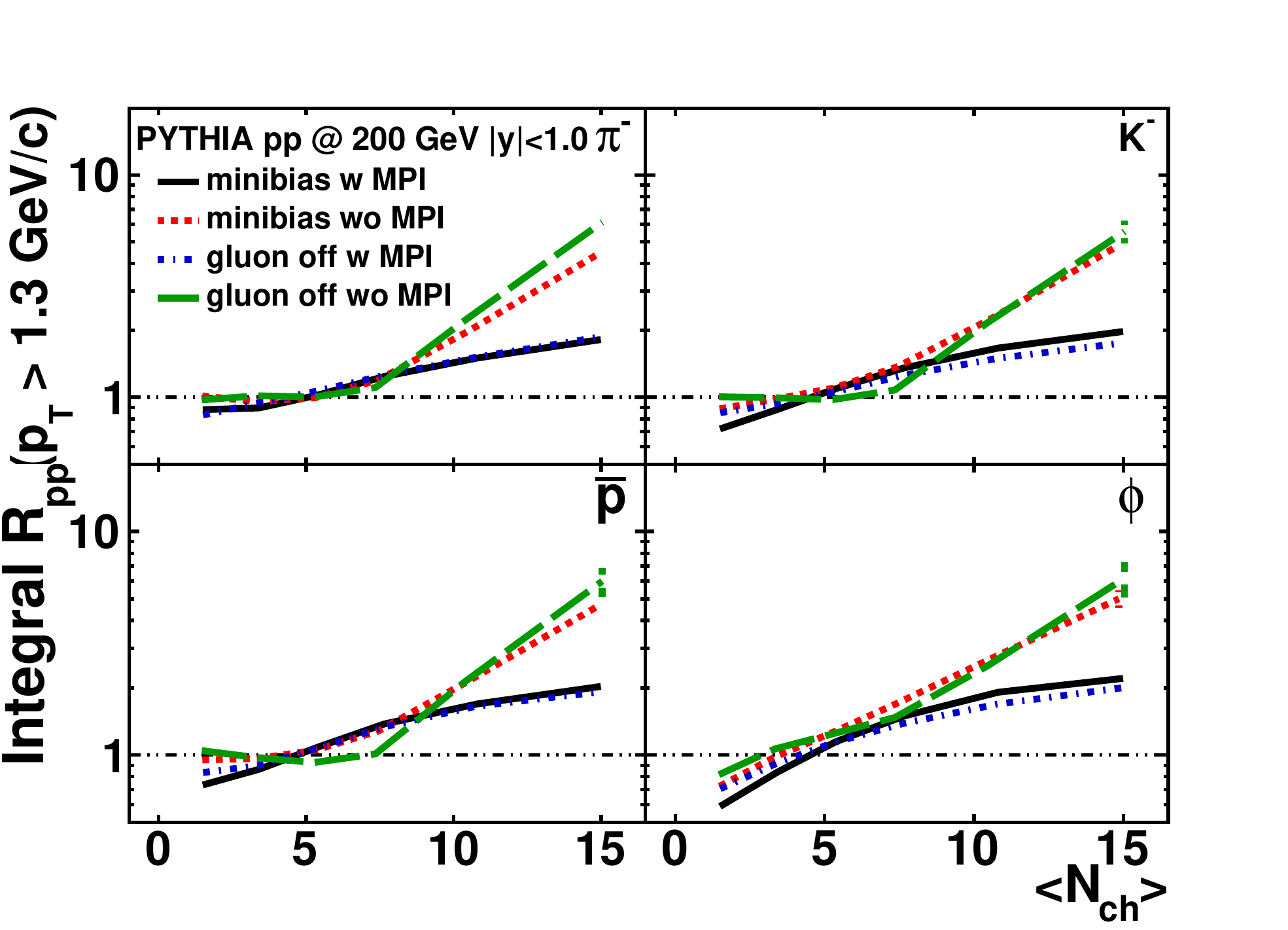}
\caption{\label{label}Integral $\rpp$ as a function of $\NchM$ for $\pi^{-}$, $K^{-}$, $\overline{p}$ and $\phi$ meson with and without parton multiple interactions and gluon contributions in p+p collisions at 200 GeV.}
\end{figure}

The multiplicity dependences of transverse momentum spectra for the $\pi^{-}$ meson with and without parton multiple interactions and gluon contributions in p+p collisions at 200 GeV within $|y| < 1.0$ are shown in  Fig. 2. The $\NchM$ in different multiplicity bins for different configurations are shown in Table I. We observed a clear hardening of the $\pT$ spectra for the $\pi^{-}$ meson from low to high multiplicity, which is consistent with the results in Ref. ~\cite{ALICEptspectrum-mult} although in different collision energies. Meanwhile the $\pT$ spectra of kaon, proton and $\phi$ meson versus multiplicity are similar as pion despite having the different integrated yields, which are not shown here.

The $\rpp$ distributions of $\pi^{\pm}$, $K^{\pm}$, $p$, $\overline{p}$ and $\phi$ meson in p+p collisions at $\spp$ = 200 GeV are extracted using the formula (1) 
The pion $\rpp$ distributions as a function of $\pT$ with the different initial production mechanisms in the different multiplicity bins are shown in Fig. 3. We observe the clear $\rpp$ splitting with different production mechanisms and the splitting is getting more obvious when $\pT$ goes higher. Parton multiple interactions suppress the splitting at higher $\pT$ range. The gluon processes contribute small in changing the relative momentum shape with respect to multiplicities. 
We find similar conclusions for the $\rpp$ distributions of kaon, proton and $\phi$ meson, which are not shown here.

The integrated $\rpp$ distributions for different particle species at $\pT > 1.3$ GeV/c are shown in Fig. 4. The increase of the integrated $\rpp$ could be due to the jet fragmentations as described in Ref. ~\cite{XINNIAN-jet}. As we can see parton multiple interaction is the dominant source of the $\rpp$ splitting suppression. This may be because of that the particle momenta get softer after multiple scatterings with energy transferring to surrounding partons. Thus the parton multiple interaction is the main competitive source to the jet fragmentations. We also find that the gluon contribution has small impact on the $\rpp$ splitting. Furthermore, qualitatively, no obvious particle-antiparticle dependence for $\rpp$ splitting is observed. More quantitive study on the production differences of particle-antiparticle and particle species is in Sect. III-C.

\subsection{\label{sec:citeref}Average Transverse Momenta $\pTM$}
In this section, the multiplicity dependence of $\pTM$ for $\pi^{-}$, $K^{-}$ and $\overline{p}$ in p+p collisions are presented and compared with that in d+Au and Au+Au collisions at $\sauau$ = 200 GeV from the STAR experiment ~\cite{STARpikpFUQIANG}. Here the charged particle rapidity density ($dN_{ch}/dy$) is used to take place of $\NchM$ in each multiplicity bin for an apple-to-apple comparison with data. It is extracted by summing up the rapidity density of pion, kaon, proton and antiproton within $|y| < 0.1 $. Fig. 5 shows the $\pTM$ distributions of $\pi^{-}$, $K^{-}$ and $\overline{p}$ as a function of $dN_{ch}/dy$ in different initial production mechanisms of p+p collisions at 200 GeV. The $\pTM$ distributions of pion, kaon and proton spectra in p+p collisions increase significantly with multiplicity, which show the same tendency as that in Au+Au collisions but with slightly larger slope. In p+p collisions, there is only hadronic processes while it is much more complicated in Au+Au collisions, in addition to jet fragmentation and parton multiple interactions, the radial flow, energy-loss mechanisms can modify the particle momentum thus changing the $\pTM$ distributions. But the similar increasing tendency as a function of charged particle density in both p+p and A+A collisions suggests that the baseline contributions of jet fragmentation and parton multiple interactions hold in different collision systems.

\begin{figure}[h]
\includegraphics[width=0.5\textwidth]{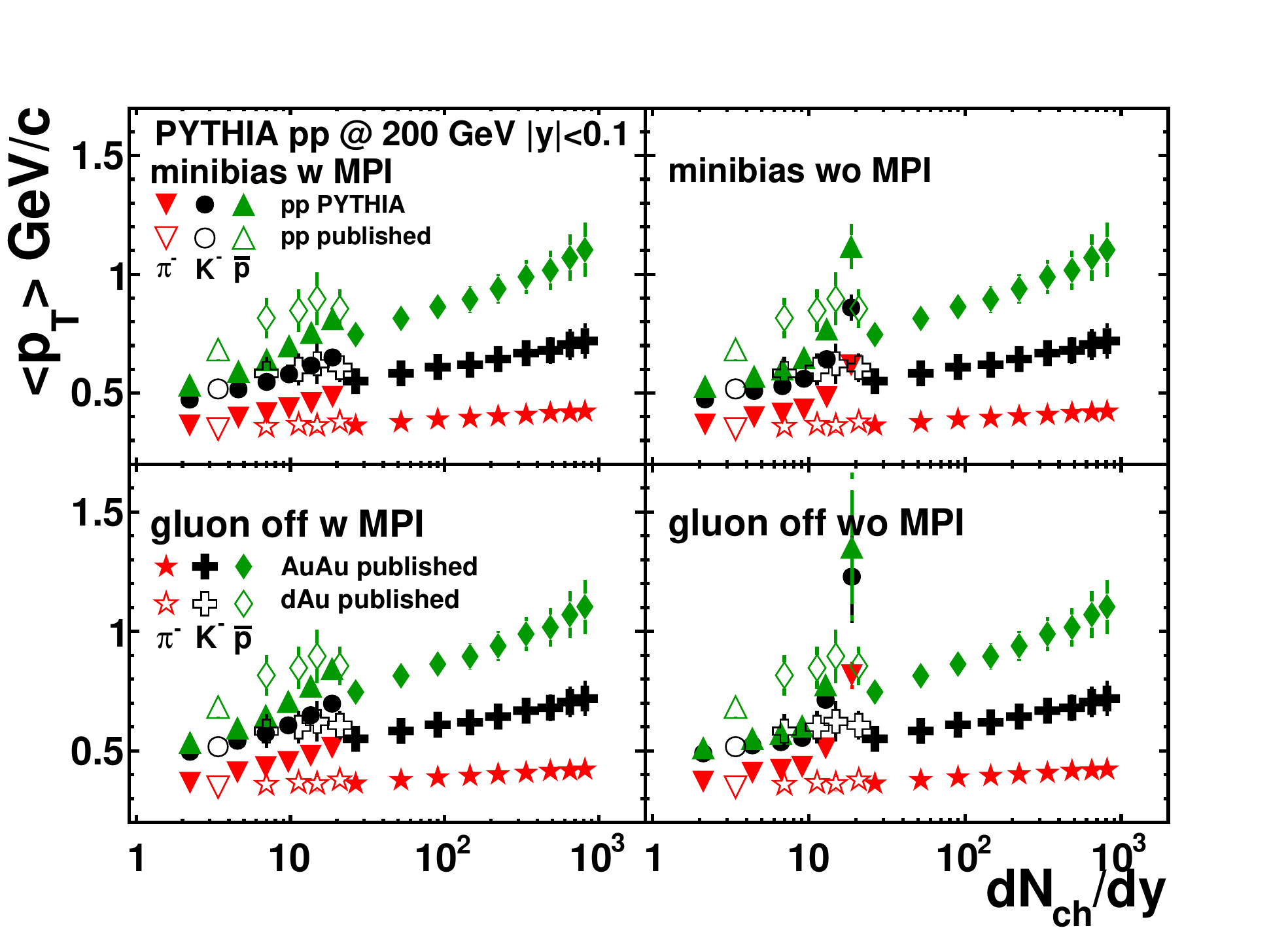}
\caption{\label{label}$\pTM$ distributions as a function of $dN_{ch}/dy$ for $\pi^{-}$, $K^{-}$ and $\overline{p}$ meson in different initial production mechanisms of p+p collisions and in Au+Au collisions at 200 GeV.}
\end{figure}

\subsection{\label{sec:citeref} Particle Ratios}
\begin{figure}[h]
\begin{minipage}{20pc}
\includegraphics[width=20pc,height=15pc]{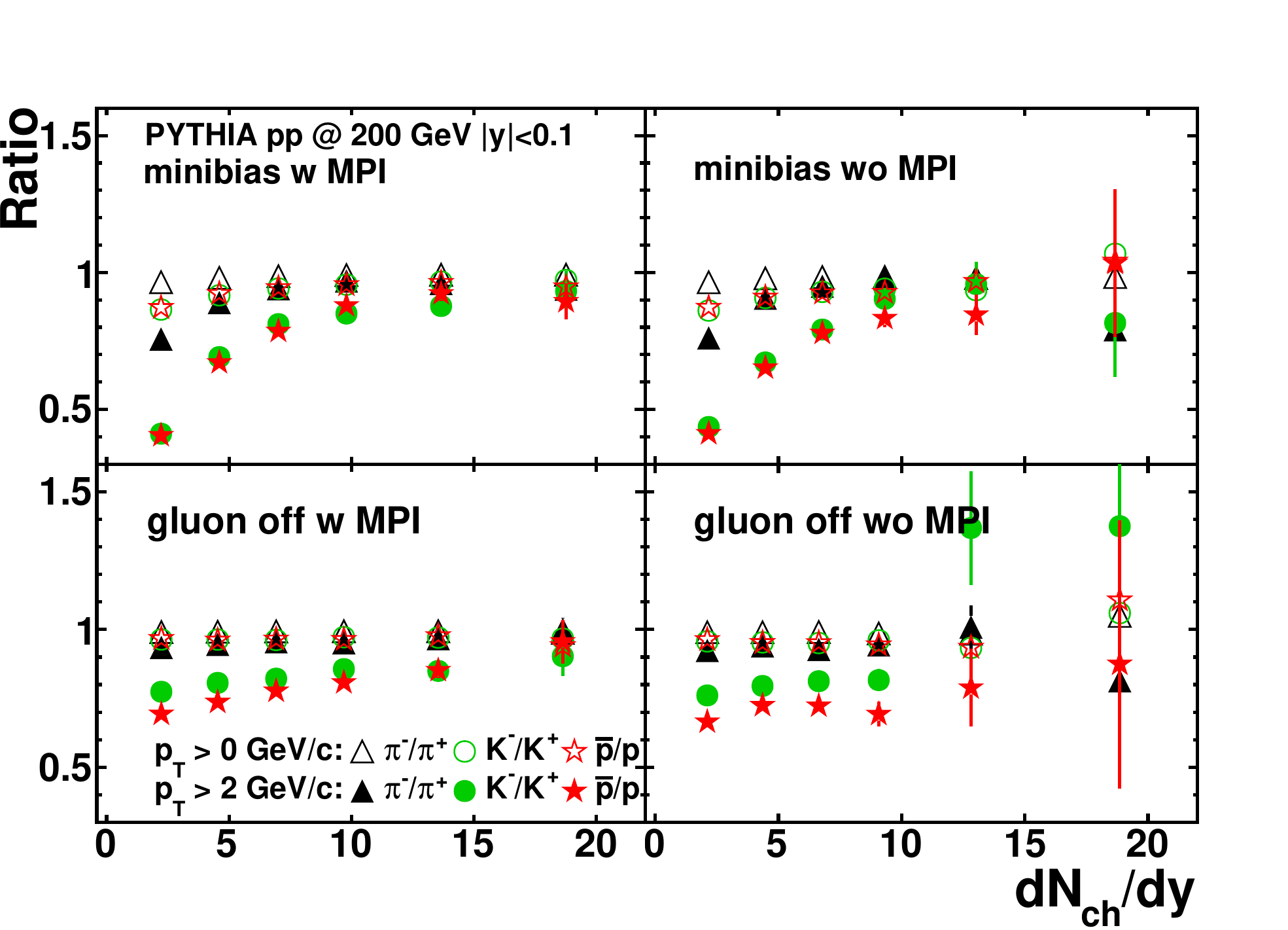}
\caption{\label{label}Ratio of antiparticle to particle as a function of $d\Nch/dy$ within $|y| < 0.1$ in different initial production mechanisms of p+p collisions at 200 GeV.}
\end{minipage}
\begin{minipage}{20pc}
\includegraphics[width=20pc,height=15pc]{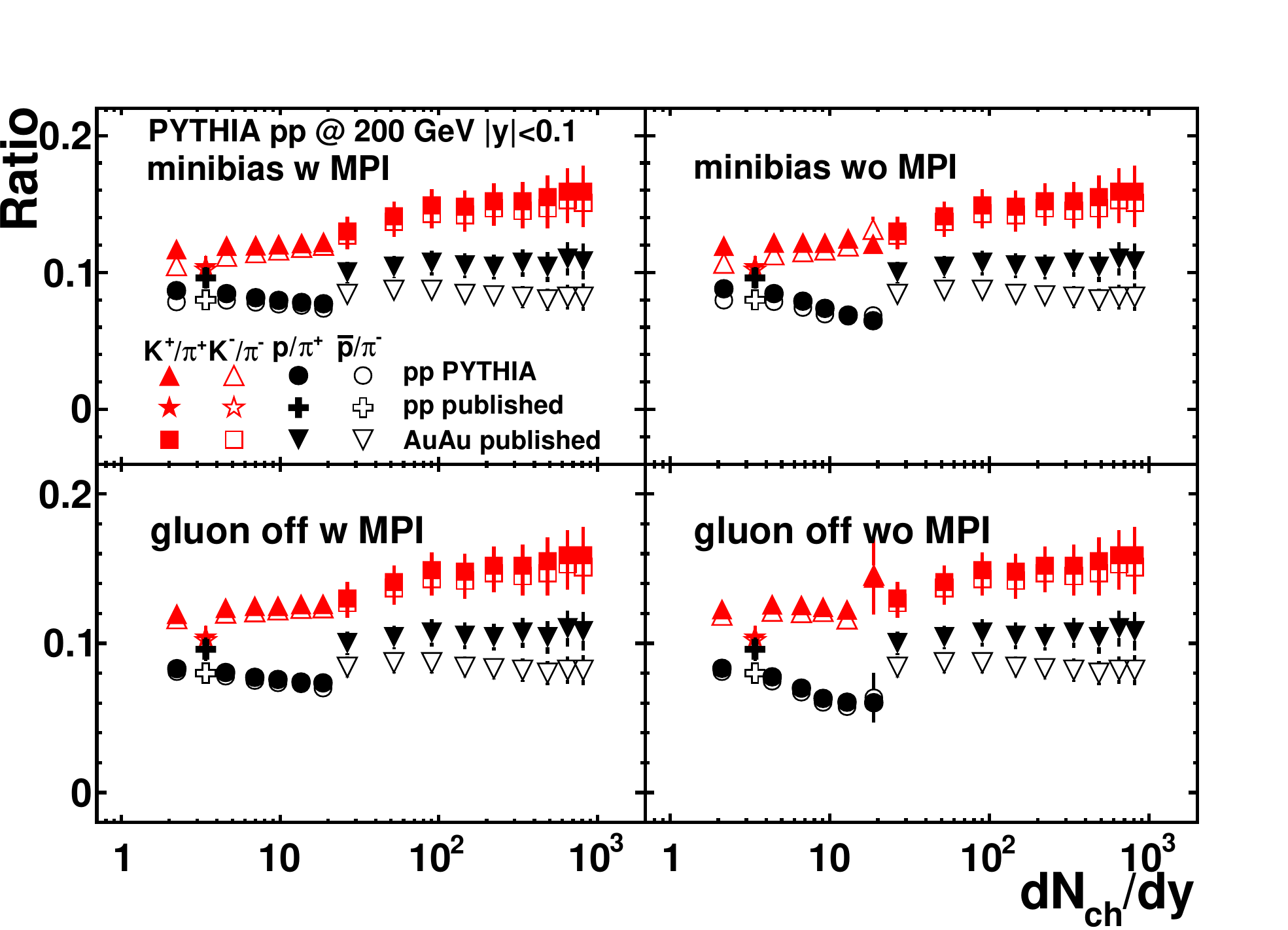}
\caption{\label{label}Ratio of $\kpipos$, $\kpineg$, $\ppipos$ and $\ppineg$ as a function of $d\Nch/dy$ within $|y| < 0.1$ in different initial production mechanisms of p+p collisions and in Au+Au collisions at 200 GeV.}
\end{minipage}
\end{figure} 
\begin{figure}[h]
\begin{minipage}{20pc}
\includegraphics[width=20pc,height=15pc]{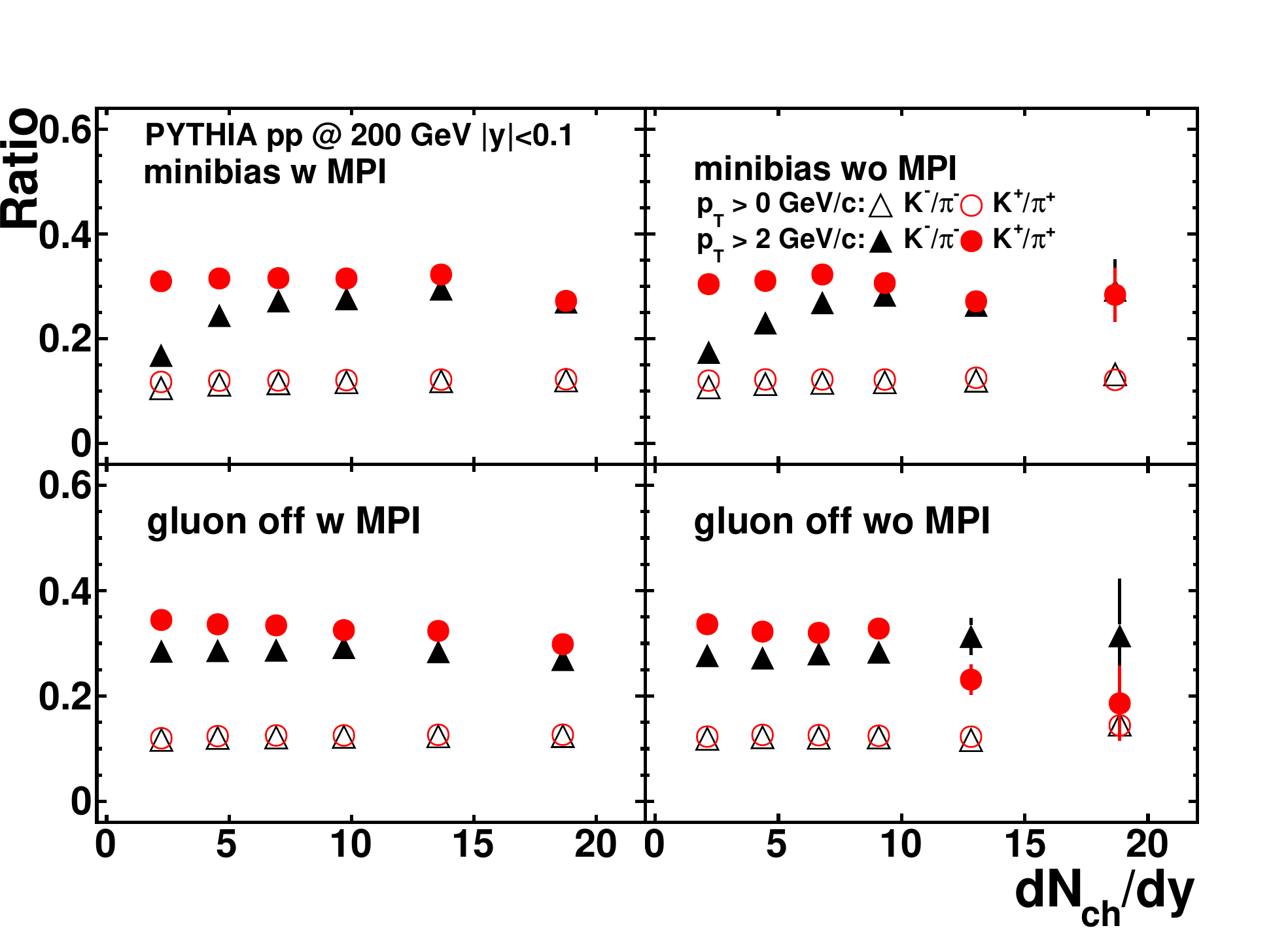}
\caption{\label{label}Ratio of $\kpipos$ and $\kpineg$ as a function of $d\Nch/dy$ within $|y| < 0.1$ at different $\pT$ regions in different initial production mechanisms of p+p collisions at 200 GeV.}
\end{minipage}
\begin{minipage}{20pc}
\includegraphics[width=20pc,height=15pc]{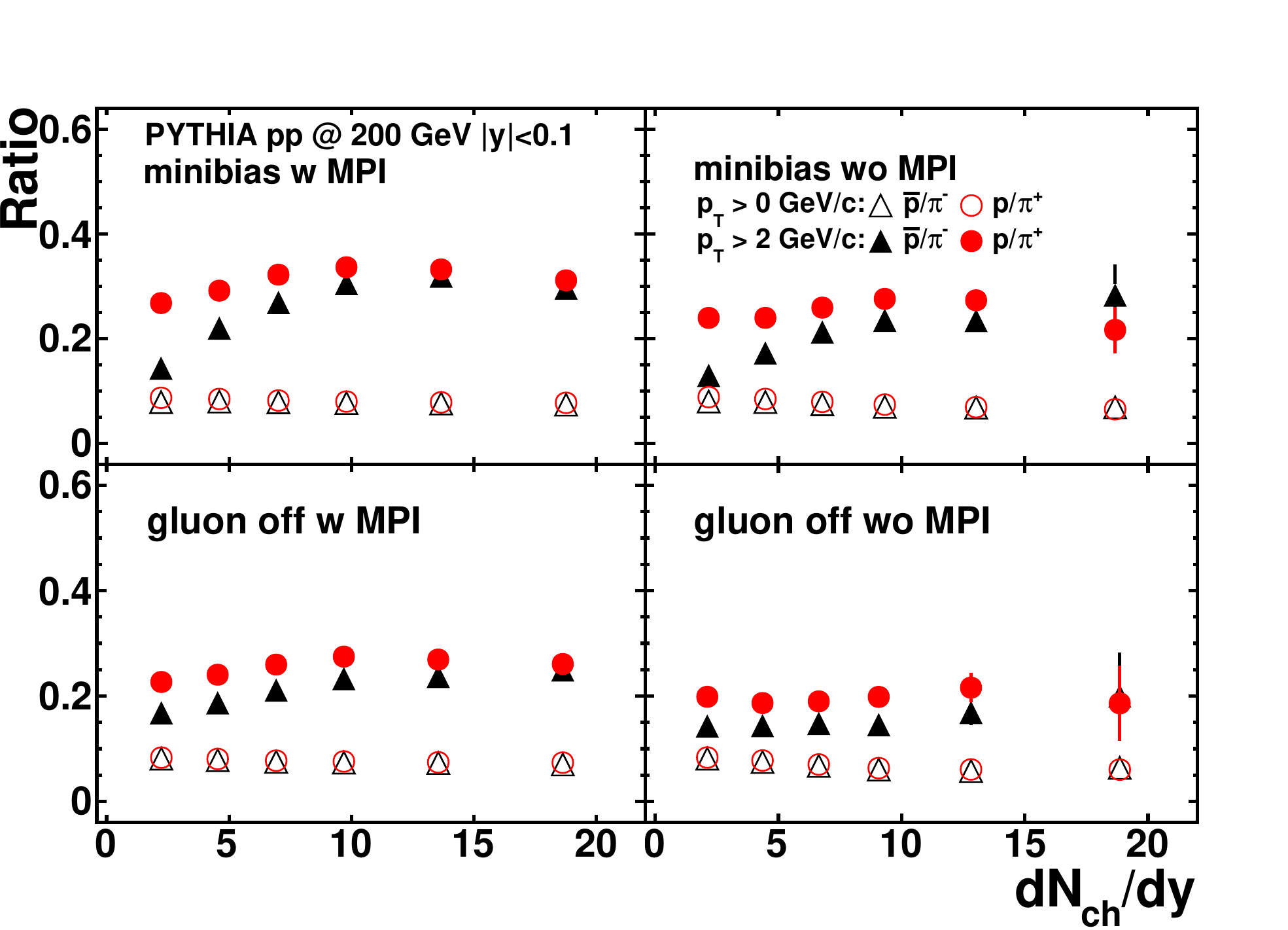}
\caption{\label{label}Ratio of $\ppipos$ and $\ppineg$ as a function of $d\Nch/dy$ within $|y| < 0.1$ at different $\pT$ regions in different initial production mechanisms of p+p collisions at 200 GeV.}
\end{minipage}
\end{figure}

In this section, the multiplicity dependence of particle ratios for $\pipi$, $\kk$, $\pp$, $\kpipos$, $\kpineg$, $\ppipos$ and $\ppineg$ in p+p collisions are presented and compared with that in Au+Au collisions. The antiparticle-to-particle ratios ($\pipi$, $\kk$, $\pp$) as a function of the charged particle multiplicity within $|y| < 0.1$ in different initial production mechanisms of p+p collisions at 200 GeV are shown in Fig. 6. In order to further understand antiparticle and particle production mechanisms, the ratio of antiparticle-to-particle was studied in different $\pT$ regions as a function of $dN_{ch}/dy$. We observe that the $\pipi$ ratio is independent of multiplicity, and the $\kk$ and $\pp$ ratios slightly depend on multiplicity in total $\pT$ regions. However, the dependence of multiplicity in small multiplicity region for $\pipi$ ratio is obvious and the dependence is stronger for $\kk$ and $\pp$ ratios in high $\pT$ region ($\pT > 2$ GeV/c). The dependence becomes smaller when we switch off the gluon contributions. 


The ratios of $\kpipos$, $\kpineg$, $\ppipos$ and $\ppineg$ for p+p collisions at $\spp$ = 200 GeV are shown in Fig. 7 for different multiplicity in different initial production mechanisms, along with a comparison to that in Au+Au collisions ~\cite{STARpikpFUQIANG, STARpikp} at $\sauau$ = 200 GeV. We also compare them with the STAR published result in minimum bias p+p collisions at the same collision energy. The ratios are consistent with the STAR published result within uncertainties in p+p collisions ~\cite{STARpikpFUQIANG, STARpikp}. The similar tendency of the particle ratios in p+p and Au+Au collisions within $|y| < 0.1$ indicates the similar production mechanisms hold in different collision systems.

Likewise, we also study the ratios of $\kpipos$, $\kpineg$, $\ppipos$ and $\ppineg$ with multiplicity in different $\pT$ regions. The results are shown in Fig. 8, 9. Weak multiplicity dependences are observed for the integrated particle yield ratios in the whole $\pT$ region, where the low $\pT$ $dN/dy$ dominates. However, in high $\pT$ region we see clear difference between particle-to-piplus and antiparticle-to-piminus ratios in low multiplicity region. This difference could be due to gluon contribution since it becomes smaller after switching off the gluon contributions. This may suggest that the high $\pT$ gluon jets compositions are different in terms of particle-to-piplus and antiparticle-to-piminus ratios. 

\begin{figure}[h]
\includegraphics[width=0.5\textwidth]{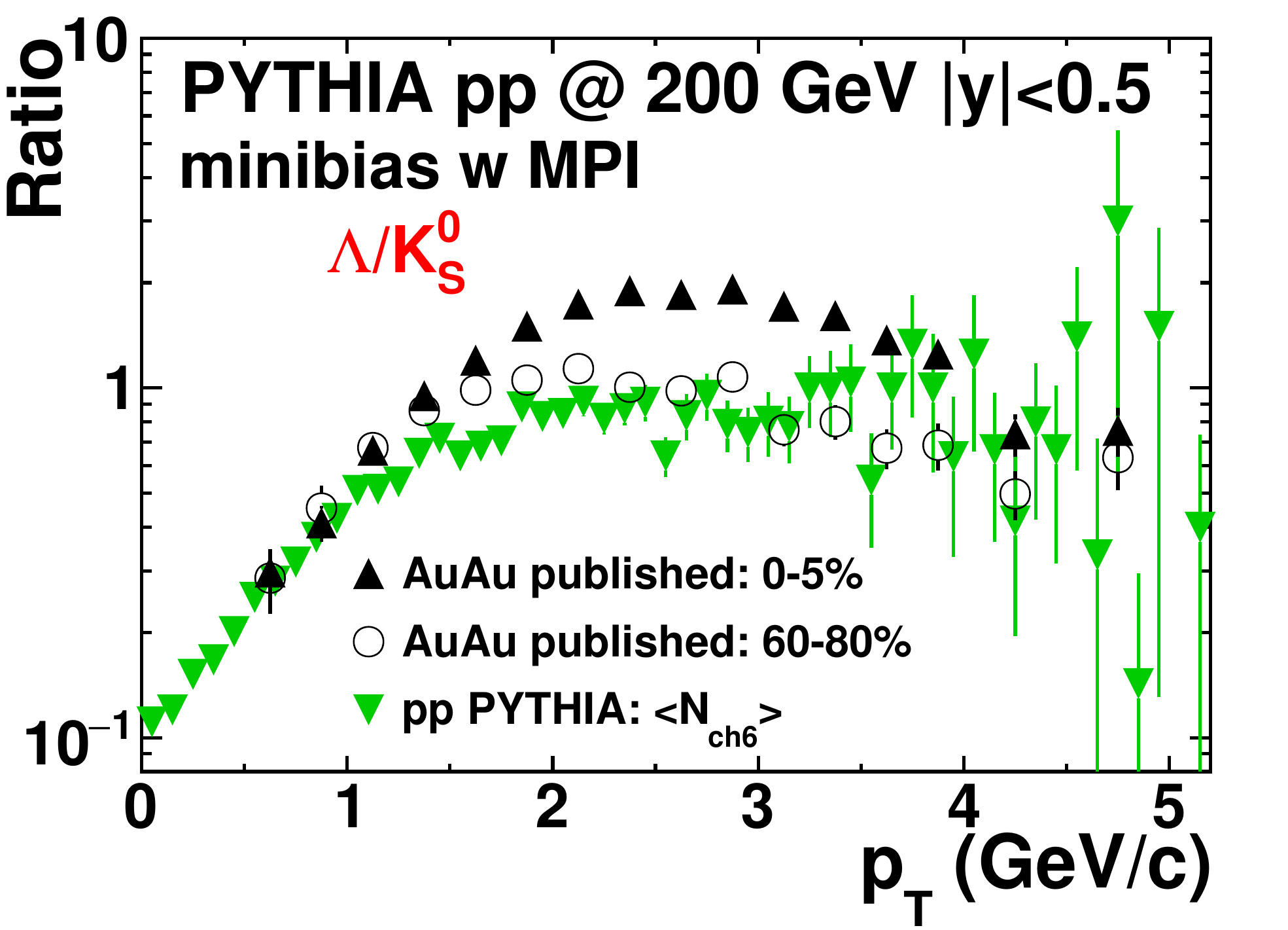}
\caption{\label{label}Ratio of $\Lambda/K^{0}_{S}$ as a function of $\pT$ with parton multiple interactions at high multiplicity in p+p collisions and in different centrality of Au+Au collisions at 200 GeV.}
\end{figure}
\begin{figure}[h]
\includegraphics[width=0.5\textwidth]{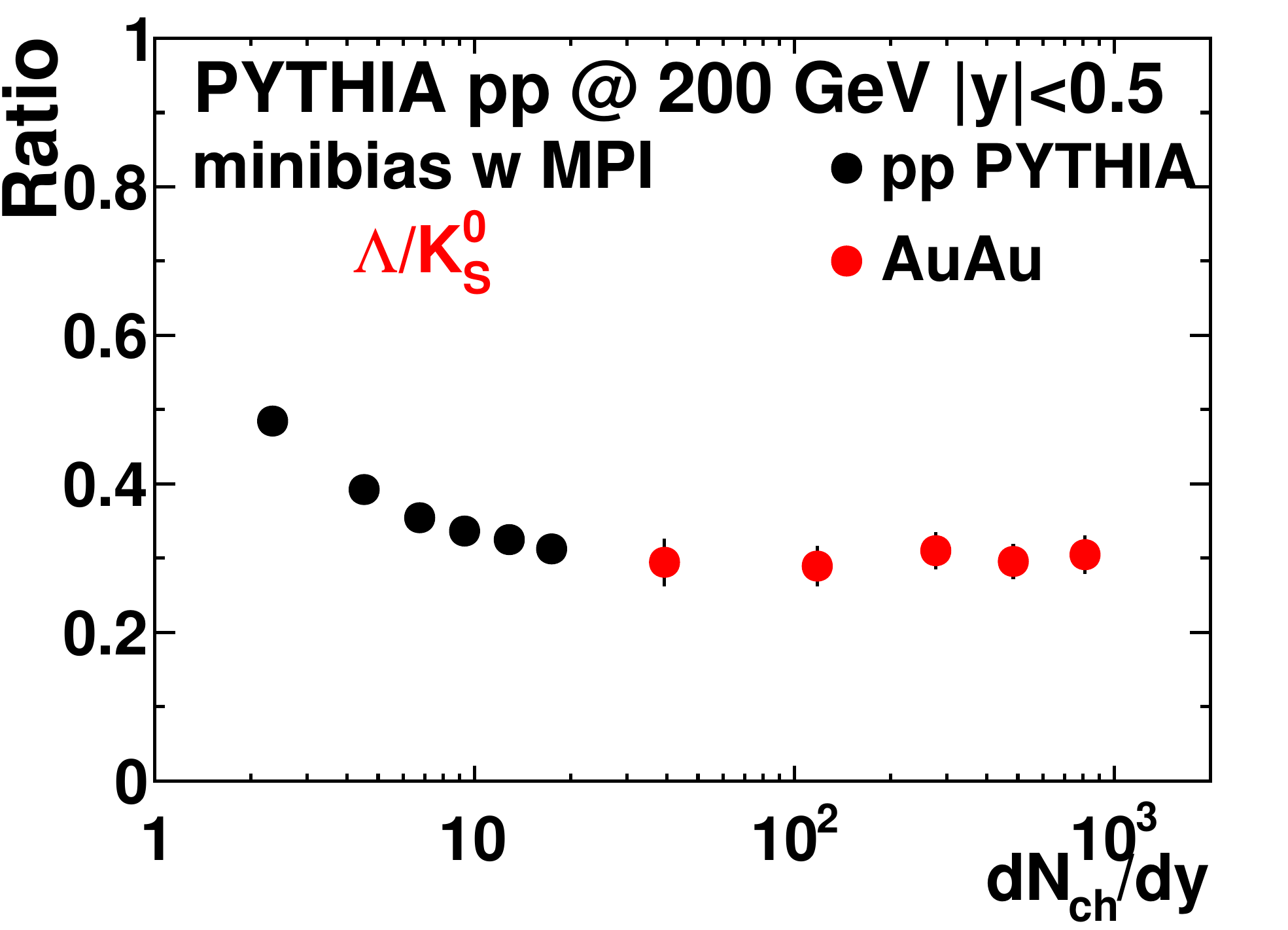}
\caption{\label{label}Ratio of $\Lambda/K^{0}_{S}$ as a function of $dN_{ch}/dy$ with parton multiple interactions in p+p collisions and in Au+Au collisions at 200 GeV.}
\end{figure}

Fig. 10 shows the ratio of $\Lambda/K^{0}_{S}$ as a function of $\pT$ at high multiplicity in p+p collisions at $\sqrt{s_{NN}}$ = 200 GeV, comparing with that in central and peripheral Au+Au collisions in the same collision energy, which is obtained by taking the ratio of the $\Lambda$ yields to $K^{0}_{S}$ yields measured at different centrality in Au+Au collisions ~\cite{STARstrangeness}. The ratio at high multiplicity in p+p collisions is consistent with data in peripheral Au+Au collisions. However, the enhancement of the $\Lks$ ratio measured in central Au+Au collisions at RHIC and p+p collisions at LHC can not be reproduced in $\pa$ simulation, where is supposed to be no hot medium created. This is a complementary support to the coalescence hadronization ~\cite{Coalescence} in nuclear medium created in Au+Au collision at RHIC or even p+p collisions in TeV at reference to ALICE nature article. 

Futhermore, the distribution of $\Lambda/K^{0}_{S}$ ratio as a function of $dN_{ch}/dy$ is studied and compared with data in Au+Au collisions. 
The $dN_{ch}/dy$ used for the Au+Au data is extracted from STAR published result by polynomial extrapolation to full the number of participating nucleons ($N_{part}$) coverage. ~\cite{STARpikpFUQIANG}. 
The distributions of $\Lambda/K^{0}_{S}$ ratio as a function of $dN_{ch}/dy$ in p+p and Au+Au collisions are shown in Fig. 11. The ratio decreases as a function of $dN_{ch}/dy$ in p+p collisions while it keeps constant in Au+Au collisions in the same collision energy. 
We found that the $\Lks$ ratio in large multiplicity events in 200 GeV p+p collisions with $\pa$ simulation is comparable with the most peripheral Au+Au collisions at RHIC. The multiplicity dependence of $\Lks$ ratio in p+p collisions shows smooth connection to that in Au+Au collisions at $\sqrt{s_{NN}}$ = 200 GeV, which may indicate some similar production mechanisms behind from transferring p+p collisions to a more hot and dense system. 

\section{\label{sec:level1}Summary}
The multiplicity dependence of charged particles of $\pi^{\pm}$, $K^{\pm}$, $p$, $\overline{p}$ and $\phi$ meson production yields at $|y| < 1.0$ in proton-proton collisions at $\spp$ = 200 GeV based on $\pa$ simulation has been presented. The $\rpp$ distributions show the obvious splitting as a function of multiplicity for all particle species, which is caused by the jet fragmentation, parton multiple interactions and gluon contributions. The parton multiple interactions suppress the splitting obviously, while the gluon contributions play less important role. The distributions of $\kpipos$, $\kpineg$, $\ppipos$ and $\ppineg$ ratios in different multiplicity in p+p collisions are compared with that in Au+Au collisions within $|y| < 0.1$, showing the similar trendency. This suggests that is similar underlying initial production mechanisms in p+p and Au+Au collisions. We also present the multiplicity dependence of the $\Lks$ ratio at $|y| < 0.5$ in p+p collisions at $\spp$ = 200 GeV based on PYTHIA simulation. The smooth connection of the $\Lks$ ratios from p+p to Au+Au collisions is observed, which may provide a hint that a smaller-sized hot medium may be created in such fundamental particle collisions with sufficient high initial energy density or some similar production mechanisms holds from transferring p+p collisions to a more hot and dense system.

\section*{Acknowledgments}
We express gratitude to Heavy Energy Nuclear Physics groups at USTC for their support. This work was supported in part by the Major State Basic Research Development Program in China with grant no. 2014CB845402, the Ministry of Science and Technology (MoST) of China under grant No. 2016YFE0104800, the National Natural Science Foundation of China with grant no. 11375184 and the Youth Innovation Promotion Association fund of CAS with grant No. CX2030040079.

\end{document}